\documentclass[twocolumns]{aa}
\usepackage[varg]{txfonts}
\usepackage[english]{babel}
\usepackage{amsmath}
\usepackage{graphicx}
\usepackage{subcaption}
\usepackage{hyperref}

\usepackage{color}
\newcommand{\refedit}[1]{#1}%{\textbf{\color{red}#1}}
\usepackage{tikz}

\begin{document}
\title{V1094 Sco: a rare giant multi-ringed disk around a T Tauri star}

\author{S.E.~van Terwisga\inst{\ref{leiden1}}\and E.F. van Dishoeck\inst{\ref{leiden1},\ref{garching}}\and M. Ansdell\inst{\ref{berkeley}}\and N. van der Marel\inst{\ref{victoria}}\and L. Testi\inst{\ref{ESO}\and\ref{INAF}}\and J.P. Williams\inst{\ref{hawaii}}\and S. Facchini\inst{\ref{garching}}\and M. Tazzari\inst{\ref{cambridge}}\and M.R. Hogerheijde\inst{\ref{leiden1}\and\ref{api}}\and L. Trapman\inst{\ref{leiden1}}\and C.F. Manara\inst{\ref{ESO}}\and A. Miotello\inst{\ref{ESO}}\and L.T. Maud\inst{\ref{leiden1}}\and D. Harsono\inst{\ref{leiden1}}}

\institute{
Leiden Observatory, Leiden University, PO Box 9513, 2300 RA Leiden, The Netherlands\label{leiden1}\\ \email{terwisga@strw.leidenuniv.nl}\and
Max-Planck-Institut f{\"u}r Extraterrestrische Physik, Gie{\ss}enbachstraße, D-85741 Garching bei M{\"u}nchen, Germany\label{garching} \and
CIPS, University of California, Berkeley, 501 Campbell Hall, CA, USA\label{berkeley} \and
Herzberg Astronomy \& Astrophysics Programs, National Research Council of Canada, 5017 West Saanich Road, Victoria, BC, Canada V9E 2E7 \label{victoria} \and
European Southern Observatory, Karl-Schwarzschild-Str. 2, D-85748 Garching bei M{\"u}nchen, German\label{ESO} \and
INAF-Osservatorio Astrofisico di Arcetri, Largo E. Fermi 5, I-50125 Firenze, Italy\label{INAF} \and
Institute for Astronomy, University of Hawai`i at M{\=a}noa, 2680 Woodlawn Dr., Honolulu, HI, USA\label{hawaii} \and
Institute of Astronomy, University of Cambridge, Madingley Road, CB3 0HA, Cambridge, UK\label{cambridge} \and
Anton Pannekoek Institute for Astronomy, University of Amsterdam, Postbus 94249, 1090 GE Amsterdam, The Netherlands\label{api}
}

\abstract{A wide variety of ring-like dust structures has been detected in protoplanetary disks, but their origin and frequency are still unclear.}
{To characterize the structure of an extended, multi-ringed disk discovered serendipitously in the ALMA Lupus disk survey and put it in the context of the Lupus disk population.}
{ALMA observations in Band 6 at 234 GHz and Band 7 at 328 GHz at 0.3$''$ resolution toward the K6 star V1094 Sco in Lupus III are presented, and its disk structure analyzed. The spectral index $\alpha_{\rm{mm}}$ is determined in the inner $150$\,AU of the disk.}
{The ALMA continuum data show a very extended disk with two gap/ring pairs. The gaps are located at 100\,AU and 170\,AU, the bright rings at 130\,AU and 220\,AU. Continuum emission is detected out to a 300\,AU distance, similar to IM Lup but a factor of 5 larger than typically found for Lupus disks at this sensitivity and resolution. The bright central region of the disk (within 35\,AU) is possibly optically thick at 1\,mm wavelengths, and has a brightness temperature of only 13\,K. The spectral index increases between the inner disk and the first ring, at the location of the first gap.}
{Due to the low temperature of the disk midplane, snow lines can be excluded as the drivers behind the ring and gap formation in this disk. Disks the size of V1094 Sco are rare, and only $2.1 \pm 1.5 \%$ of disks in Lupus show continuum emission beyond 200\,AU. Possible connections between the large primordial disk population, transition disks, and exoplanets are discussed.}


\maketitle

\section{Introduction}
	One of the key discoveries of the Atacama Large
Millimeter/submillimeter Array (ALMA) has been the unexpectedly wide
variety of structures seen in protoplanetary disks, ranging from
(asymmetric) dust cavities to multiple rings at a range of
radii. These multiple rings have been found in disks around both Herbig Ae/Be and
T Tauri stars~(\citealt{isella16, fedele17a, walsh16, vdplas17,
  walsh14,hltau, twhya, loomis17, cieza17, dipierro18, fedele17b}). Of these T
Tauri stars, HL Tau, Elias 2-24, and AS 209 are very young
objects ($< 1$\,Myr), while the massive TW Hya disk is unexpectedly
old (up to 10 Myr). The bulk of the disks imaged in continuum surveys
of nearby star-forming regions such as Lupus does not seem to show
substructure at 0.2 -- 0.3$''$ resolution, apart from transition disks with large inner dust cavities ($> 20$\,AU)~\citep{ansdell16,tazzari17,vandermarel18,ansdell18}.
	
	The existence of multi-ringed disks raises important questions
on the evolution of disks, and therefore on planet formation, since dust
particles trapped in a ring may grow efficiently without being
accreted onto the star~\citep[e.g.,][]{pinilla12}. Thus, rings can
provide the material necessary for possible future planet
formation. So far, several mechanisms have been proposed in the
literature to generate ring-like continuum structures:
(secular) gravitational instabilities, the effect of snow lines of
various molecules, dead zones and, most tantalizingly, the presence of
already-formed planets in the intervening gaps~\citep[e.g.,][]{takahashi14, zhang15,
  okuzumi16, flock15, papaloizou84}. Our understanding of the physics
behind these multi-ringed disks is limited by the small number of known
examples ($N = 5$ around T Tauri stars), as well as the difficulty of
placing them in the context of a well-studied, similar population.
Moreover, it is still not clear how common it is for disks to show
multiple continuum rings, whether they correlate with stellar or disk
parameters, or when in their evolution features like these are most
likely to be present.

	Another, less obvious structure found in many protoplanetary disks is an intensity profile with a bright central region surrounded by a fainter, extended outer region, resembling a ``fried egg''. This structure can be seen both in multi-ringed disks, such as AS 209 and HD 163296~(\citealt{isella16, fedele17b}), and in full disks where no rings have been identified, such as V883 Ori~\citep{cieza16}. In the well-studied case of V883
Ori, the origin of the morphology is due to a steep (spatially
unresolved) increase in optical depth at the location of the knee in
the intensity profile. This increase may be linked to the presence of
a water snowline, inside of which the originally large icy grains
fragment into many smaller bare silicate grains. These smaller
particles have smaller drift velocities causing a 'traffic jam' and
thus a dust pile-up in the inner disk increasing the optical depth
\citep{birnstiel10,pinilla15,banzatti15,cieza16}. More generally, this
type of intensity profile must trace the underlying grain properties
(number density, composition and size, or both).
	
	In this paper, we present ALMA Band 6 and Band 7 observations of V1094
Sco, a newly-identified giant multi-ringed disk in Lupus. V1094 is a K6
star with a luminosity of 1.7\,(1.95) L$_{\odot}$ and an accretion rate of $\log( \dot{M}_{\rm{acc}} / (M_{\odot} \rm{yr}^{-1}) ) = -8.12\,(-7.7)$~\citep{alcala17}. Throughout this article, we take a 150\,pc distance to the source, consistent with the latest {\it Gaia} distance estimate of $153 \pm 1$\,pc~\citep{gaiadr2, bailerjones18}; the values between brackets correspond to the published values assuming a pre-{\it Gaia} 200\,pc source distance.
%%% NB: The following sentence was in the previous version; it has been superseded by the Gaia DR 2 results (these do not affect our conclusions, but we felt it would be more accurate.)
% It is part of the Lupus III complex, taken here to have a distance of 150 pc based on three {\it Gaia} parallaxes of confirmed Lupus III members~\citep{gaiadr1, long17}; the luminosity and mass accretion rates between brackets correspond to the published values assuming a pre-{\it Gaia} 200\,pc source distance.
The V1094 disk had previously been recognized as an enigmatic object,
in particular by~\citet{tsukagoshi11}, who inferred from spectral
energy distribution (SED) modeling that it was a cold and massive
disk.  Photometry from the {\it Herschel Space Observatory} shows a
large far-infrared excess, with an SED classification inconsistent
with that of a transition disk \citep{bustamante15}. Previous direct imaging searches
for companions, either stellar or of planetary mass, have not found
any down to $\sim 10\,M_J$ beyond 75\,AU~\citep{joergens01,uyama17}.
		
	Several key aspects of this disk make it an especially valuable
addition to the so far small sample of multi-ringed disks around T
Tauri stars: first, the population of Class II Young Stellar Objects
in the Lupus clouds is exceptionally well-studied, with a $100\%$
coverage in ALMA Band 6 and 7 continuum at $0.3''$ resolution and
$0.3$\,mJy beam$^{-1}$ sensitivity \citep{ansdell16,ansdell18}. $80\%$ of these sources, including all the objects discussed here, also have VLT/X-Shooter
spectroscopy data available resulting in accurate stellar types and
accretion properties \citep{alcala14,alcala17}. This enables us to compare the
disk and its host star to a large, homogenous sample, and calculate
the occurrence rate of this type of object at the age of the Lupus
cloud. Moreover, at $\sim 3$\,Myr, the age of this disk places it in
the middle between the young ($<1$\,Myr) objects, such as Elias 2-24
and HL Tau, and $10$\,Myr-old TW Hya disk~\citep{wichmann97,sokal17}.

	In \S 2, the ALMA observational details are discussed, whereas
continuum images showing a bright central millimeter emission core
with two outer gaps and rings are presented and analyzed in \S 3. The
availability of data at two frequencies allow the spectral index as
function of radius to be determined. The V1094 Sco disk is compared with
other similar disk structures in \S 4.1, and put in the context of the
full Lupus disk sample in \S 4.2, demonstrating that such large disks
are rare.

\section{Observations and data reduction}
\label{sec:obsred}
	
	The data presented here are part of the Lupus completion survey (ALMA
proposal 2016.1.01239.S, P.I.: S.E. van Terwisga) which aimed to
provide data on the sources not previously included in the ALMA survey
of Lupus protoplanetary disks \citep{ansdell16,ansdell18} and for which no data at
similar spectral and resolution settings were available at the
time. The sources included are EX Lup, GQ Lup, RX J1556.1-3655, Sz 102, Sz 76, Sz 77, and V1094 Sco.

	Our observations cover Band 6 and Band 7 continuum, and the $^{12}$CO $2-1$, $^{13}$CO $3-2$ and $2-1$, and C$^{18}$O $3-2$ and $2-1$ lines, as well as CN $3-2$. The Band 6 observations were performed on 7 July 2017, using 44 antennae and $2600 - 16700$\,m baselines, and a PWV of around 0.64\,mm. The sample was observed in Band 7 on 21 May 2017, using 41 antennae and baselines between 1100\,m and 15100\,m at PWV columns of $\sim$ 0.7\,mm. On-source integration times were 3 and 4.5 min in Band 6 and 7, respectively. 

	The Band 6 data were flux-calibrated using J1427-4206, and had J1610-3958 as phase- and J1517-2422 as bandpass calibrators. The Band 7 data used J1517-2422 as flux- and bandpass calibrator and J1607-3331 and J1610-3958 as phase calibrator. While no additional flux calibration on Solar-system objects was done, the Band 6 data were observed on the same day as their flux calibrator, and therefore have a small absolute flux error ($\sim 10 \%$) typical of ALMA. However, the Band 7 flux calibrator was both highly variable (by a factor of 1.5 in the month following the observations) and not observed close to the date on which our data were obtained. As such, its flux --- and therefore the flux calibration of all Band 7 data --- is very uncertain, complicating the calculation of spectral indices and disk masses. 
	
	Fortunately, the Band 7 data in this proposal also cover GQ Lup, which has previously been observed in that band at high $S/N$ and with excellent flux calibration (on Titan and Ceres) by~\citet{macgregor17}. In addition to their accurate fluxes (a less than $10\%$ systematic uncertainty is quoted), these authors do not find any evidence for variability of this source at mm-wavelengths, retrieving fluxes consistent with previous studies in the $1997 - 2015$ period~(\citealt{nuernberger97, dai10}). Because our observations covered GQ Lup and V1094 Sco on the same day, it is safe to compare these data with the~\citet{macgregor17} results to re-calibrate the Band 7 fluxes. This gives a flux ratio of $F_{\rm{MacGregor\,et\,al.}} / F_{\rm{this\,work}} = 1.3 \pm 0.009$, which we will implicitly apply to the Band 7 data throughout this paper.
	
	Phase-only self-calibration was used to maximize the $S/N$ of the observations in both bands. The Band 6 data allowed self-calibration down to the integration interval ($6$\,s), whereas for Band 7 we found useable solutions down to a $15$\,s interval.
		
	Imaging the continuum data using the \texttt{CASA}~\texttt{tclean} task with Briggs weighting with a robust parameter of 0.5, the beam shapes become $0.25'' \times 0.22''$ and $0.19'' \times 0.18''$ in Band 6 and Band 7 with rms noise values of 0.11 and 0.3 mJy\,beam$^{-1}$ respectively for V1094 Sco. 

	\refedit{The discussion in this article will focus on the high-$S/N$ continuum observations. Fluxes for the CO (isotopologue) lines can be found in~\citet{ansdell18}, and CN is discussed in Van Terwisga et al. (in prep.). The gas emission covers a (very) extended region, so that it is partly resolved-out. Also, the $S/N$ of the line data is low, with just a few minutes of integration time, so they are not useful for a spatially resolved analysis.}

\section{Results and analysis}
\begin{figure*}[!htb]
\begin{center}
		\includegraphics[width=1.0\textwidth]{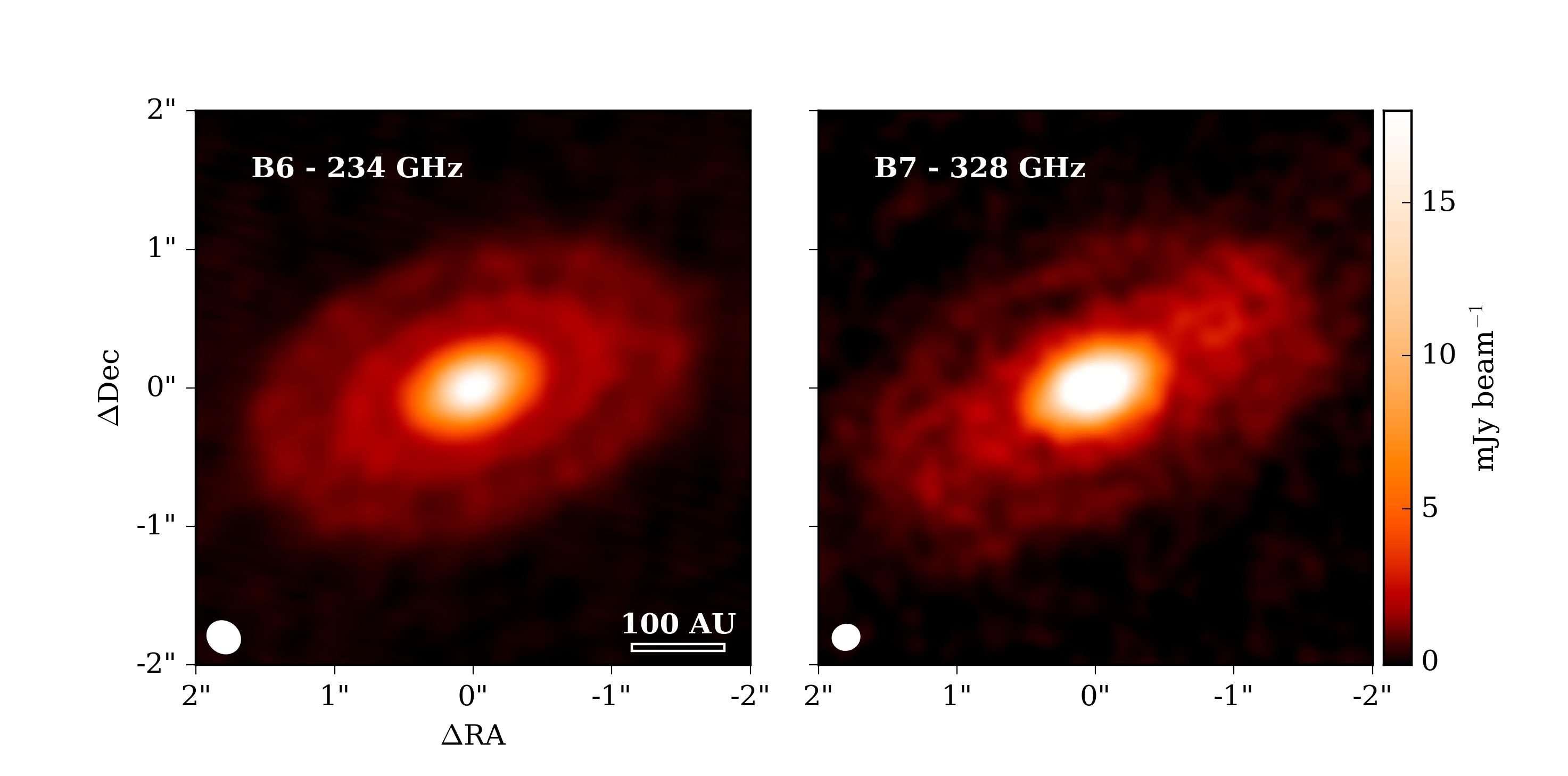}
		\caption{Continuum emission of the V1094 Sco disk in
                  ALMA Band 6 (left panel) and Band 7 (right panel). Peak fluxes are
									19.5\,mJy\,beam$^{-1}$ (Band 6) and 28.6\,mJy\,beam$^{-1}$ (Band 7) within the
									central beam.	Beam sizes are indicated in the bottom right; the physical distance scale
                  assumes a 150 pc distance towards the source.}
		\label{fig:contmap}
\end{center}
\end{figure*}

	Figure~\ref{fig:contmap} presents the Band 6 and Band 7 continuum
images resulting from the data reduction procedure described
above. The continuum rings and bright central core are immediately
obvious, especially in the Band 6 image. Also evident is the large
radial extent of the source, with continuum emission detected up to 300\,AU when placed at 150\,pc. This is comparable to the radius of the well-known massive IM Lup disk~\citep{cleeves16}, one of the largest and most massive disks in Lupus. The full integrated flux (inside of a $2.16''$ radius aperture, beyond which the image is entirely dominated by noise) in Band 6 is $180 \pm 14$\,mJy, while in Band 7 it is \refedit{$390 \pm 19$\,mJy. The fluxes and errors here were derived using aperture photometry with a curve-of-growth method to set the aperture radius, identical to the method used in~\citet{ansdell16,ansdell18}.}

	The key features are clearly the same in both images: the disk has a large radial extent, with two bright rings superimposed over a fainter, extended continuum, and a bright and at this resolution apparently featureless central region. Like other multi-ringed disks, V1094 Sco does not show significant azimuthal asymmetries at this sensitivity and resolution. Assuming the disk is indeed purely radially symmetric, fitting a two-dimensional Gaussian in the image plane with the~\texttt{imfit} task in CASA led to a beam-deconvolved PA$= 109 \pm 1.9 ^{\circ}$ and $i = 53 \pm 1.3 ^{\circ}$, and a J2000 phase center of 16h08m36.17s, -39$^{\circ}$23m02.87s.

Comparing the Band 6 and Band 7 data in the image plane, however is not entirely straightforward. Due to the lower $S/N$ and worse weather conditions affecting the Band 7 data, the phase-calibration is only partially successful. As a result, the Band 7 phase noise is higher than that of the Band 6 observations.~\refedit{Also, the $u,v$-plane coverage of the Band 7 data is different. These conditions particularly affect} the emission in the North-East and South-West of the disk~\refedit{at $\sim 1''$ from the disk center}, and makes the presence of the rings less obvious. The Band 6 data, on the other hand suffer from a low-amplitude North-South ripple, since the longest baselines are only sampled in a single direction. However, its impact is relatively minimal. Therefore we choose the Band 6 data as the basis for the subsequent analysis.
	
	\subsection{Disk structure: core, gaps and rings}
	\label{sec:uvplane}

	In order to compare the continuum structure of V1094 Sco with that of other, similar disks around both earlier and similar spectral-type stars, we model the intensity profile using an approach similar to that of~\citet{zhang16}: the intensity at radius $R$ is considered as the sum of Gaussian contributions, multiplied by a cosine.~\refedit{This model is then Fourier-transformed and fit to the derotated, deprojected data in the $u,v$-plane, after binning the data to $10$\,k$\lambda$ bins.} The advantage of this method is that it uses a minimal number of parameters and can easily formally confirm the presence of structures, and precise locations of the brightness maxima and minima. Moreover, it does not rely on a model-dependent temperature profile or assumptions about the dust properties. The $u,v$-plane data show four local maxima between 0 and 700\,k$\lambda$; therefore, we use up to $n=4$ terms, for a total of 13 free parameters.

\begin{equation}
\label{eq:zhangmodel}
\begin{split}
	I(\theta) = & \frac{1}{\sqrt{2 \pi} \sigma_0} \exp \left( - \frac{\theta^2}{2\sigma_0^2} \right) \\
	& + \sum_1^3 \cos (2\pi\theta\rho_j) \times \frac{a_j}{\sqrt{2\pi}\sigma_j} \exp \left( -\frac{\theta^2}{2\sigma_j^2} \right).
\end{split}
\end{equation}
	
	In this model, $\theta$ is the angular separation from the source center, $\sigma_i$ the width of the $i$th component, $a_i$ its intensity and $\rho_i$ the spatial frequency of the cosinusoidal perturbation. 
No baselines longer than $700$\,k$\lambda$ are included in the fit, since the data are noise-dominated beyond that point. A Monte Carlo Markov Chain (MCMC) fitting procedure was used to infer the best fit parameters, using the \texttt{pymc} package. Since the model is degenerate if the ranges for $\rho_i$ overlap, their allowed values were restricted based on the location of the peaks in the u,v-plane, but always at least $100$\,k$\lambda$ wide~\refedit{in order to sample the full width of each peak}. We confirmed that the behavior of the Markov chain was proper, \refedit{by ensuring the walkers were burnt-in and inspecting their sampling of the parameter space, and checking that the posterior distributions for the parameters showed a single clear maximum}. Since we start our model with values for the disk inclination, position angle, and coordinates, derived with \refedit{the}~\texttt{imfit} task, the resulting best-fit parameters are conditional on these priors. \refedit{We confirmed their impact on the resulting fit was minimal by re-running the fitting code with different initial choices of inclination, position angle, and source center within their error bars.}

\begin{figure}[!htb]
\begin{center}
		\includegraphics[width=0.5\textwidth]{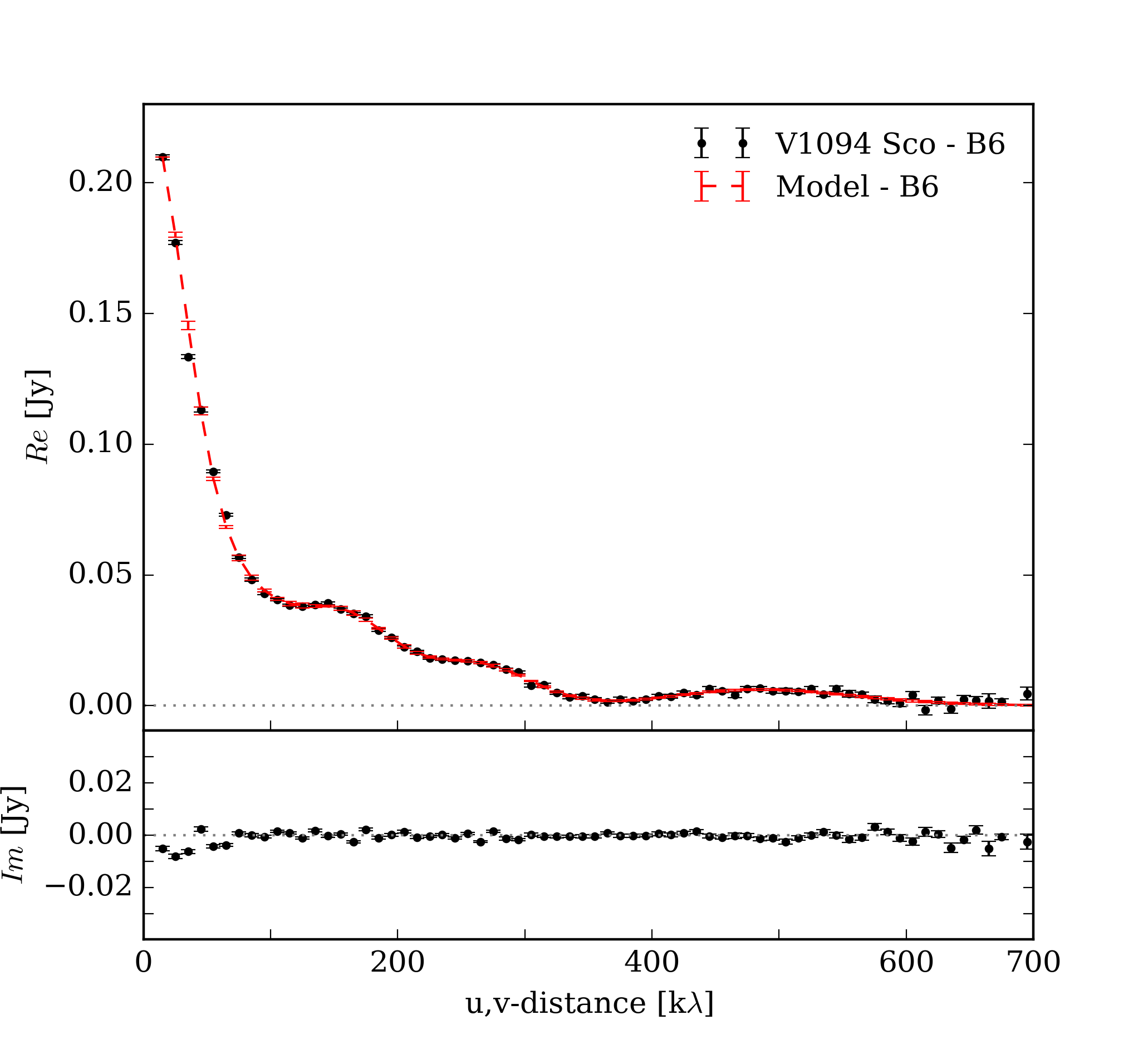}
		\caption{The best-fit model, presented in Table~\ref{tab:uvfit} (red) and deprojected, derotated and binned $u,v$-plane data for V1094 Sco (black).}
		\label{fig:uvfit}
\end{center}
\end{figure}

\begin{figure}[!htb]
\begin{center}
		\includegraphics[width=0.5\textwidth]{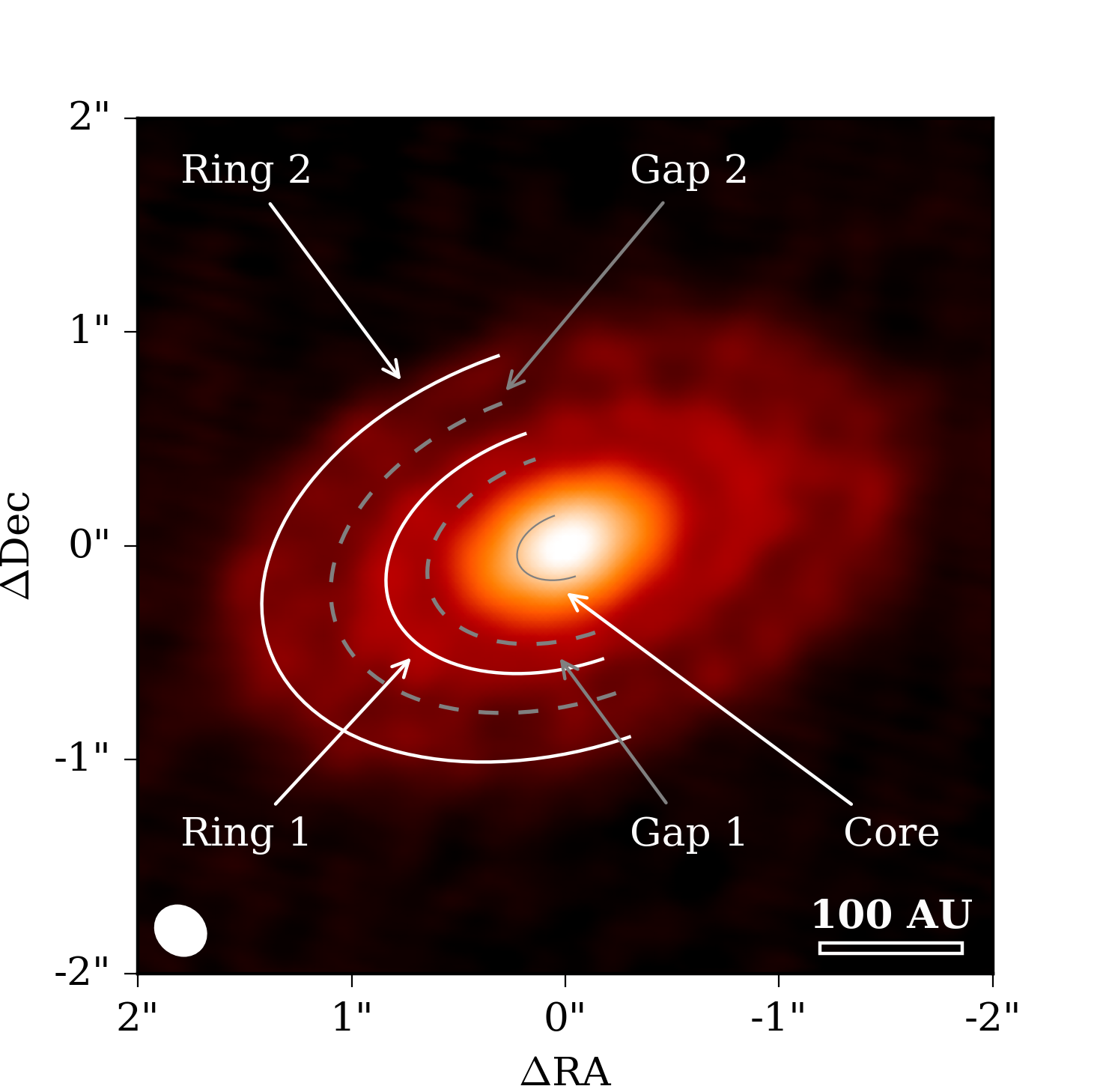}
		\caption{Best-fit locations of the bright rings (white) and gaps (gray) and the bright core ($\sigma_1$ in the model) as listed in Table~\ref{tab:imres}, superimposed on the Band 6 continuum.}
		\label{fig:annotated}
\end{center}
\end{figure}

	Figure~\ref{fig:uvfit} presents the resulting model of the intensity distribution for the Band 6 data, with the best-fit values in Table~\ref{tab:uvfit} and inferred positions for the gaps and rings in Table~\ref{tab:imres}. The best-fit ring and gap positions are also shown on top of the Band 6 data in Figure~\ref{fig:annotated}. The results show that our 4-component model is a good description of the data, fitting all local maxima within this baseline range. The parameter values also indicate our description of the disk, as a bright core with an extended outer region featuring two rings, is accurate: the `core' of the disk is contained within the $i=1$ component, having a characteristic radius of $35 \pm 2$\,AU. 
	
	In the best-fit model, the two bright rings and two gaps are also present in the image plane, arising from the combination of the various components in Equation~\ref{eq:zhangmodel}. They are identified by taking local maxima and minima of $I(\theta)$. Taking into account the errors on the fit parameters, the locations of the bright rings become $130 \pm 9$\,AU and $220 \pm 16$\,AU, while the gaps are found at $100 \pm 6$\,AU and $170 \pm 13$\,AU. This closely matches the intensity profile of the disk after deprojecting and derotating in the image plane (see also \S~\ref{sec:ringcomp}). The best-fit model gives an intensity ratio of $2.3$ for the first ring/gap combination and $1.4$ for the second ring and gap. It is difficult to constrain the width of the gaps and rings from this fit, as there may be structure at smaller scales that is not probed, due to insufficient $S/N$ at large baselines and the resolution limit.

\begin{table*}[!ht]
\caption{Parameter values and inferred errors from the MCMC fit to the Band 6 $u,v$-data.}
\label{tab:uvfit}

	\centering
	\begin{tabular}{l l l l l l}
	\hline \hline
	                      & $i=0$       & $i=1$        & $i=2$          & $i=3$         & $i=4$          \\
	\hline
	$a_i$ [Jy]            &  -          &$1.8 \pm 0.3$ &$0.64 \pm 0.096$&$0.46 \pm 0.05$&$0.52 \pm 0.18$ \\
	$\sigma_i$ [AU]       &$161 \pm 8.3$&$35 \pm 1.9$  &$57 \pm 5.0$    &$130 \pm 37$   &$120 \pm 29$    \\
	$\rho_i$ [k$\lambda$] &  -          &$20.7 \pm 0.1$&$446 \pm 12$    &$146 \pm 15$   &$255 \pm 9$     \\
	$ \lambda / \rho_i $ [AU] & -       &$1495 \pm 7$  &$69\pm1.9$      &$212\pm22$     &$121\pm4$       \\
	\hline
	\end{tabular}
\end{table*}

\begin{table*}[!ht]
\caption{Ring and gap positions from the MCMC fit to the Band 6 $u,v$-data.}
\label{tab:imres}

	\centering
	\begin{tabular}{l l l l l l}
	\hline \hline
		     &  Gap 1  &  Ring 1 &   Gap 2  & Ring 2   \\
	\hline
	R [AU] &$100\pm6$&$130\pm9$&$170\pm13$&$220\pm16$\\
	\hline
	\end{tabular}
\end{table*}	
	
	\subsection{Spectral index analysis}
	\label{sec:alpha}
	The continuum observations of V1094 Sco show that in both Band 6 and Band 7, the structure is dominated by two azimuthally symmetric regions: the bright, inner `core' and a fainter, more extended outer disk, where the rings are located. Directly linking intensity to dust properties is difficult, but necessary for tying these observations to models of ring formation. Thanks to the range in frequencies covered by our observations (between 234 and 328 GHz), the spectral index $\alpha_{\rm mm}$ of the continuum emission can be determined. In particular, if the intensity at a given radius is given by $I_{\nu} = B_{\nu}(T)(1 - e^{-\tau_{\nu}})$ and $\tau \sim \nu^\beta$, then if $\tau \gg 1$ and $h\nu / k_B T \ll 1$, $\alpha_{\rm mm} = 2$, while if $\tau \ll 1$, $\alpha_{\rm mm} = 2 + \beta$.~\refedit{However, these simple relations only hold in the Rayleigh-Jeans regime. If this does not apply, an extra (negative) temperature-dependent term appears, and values of $\alpha_{\rm{mm}}$ lower than 2 can occur.}
	
	The resolution and sensitivity of our data allow us to determine the spectral index as a function of radius. However, we cannot immediately use the results presented in Figure~\ref{fig:contmap}, since their imaging beam sizes and the $u,v$-plane coverage of the underlying data differ. To ensure that $\alpha_{\rm{mm}}$ contains only variations due to actual spectral index variations and image noise, the data were re-imaged with a circular, $0.25''$-radius beam, using only baselines with lengths between $17 - 1260$\,k$\lambda$.~\refedit{This is the largest baseline range with which it is possible to cover all the continuum spectral windows in both Band 6 and Band 7.} The profile of $\alpha_{\rm{mm}}$ was subsequently derived by deprojecting and derotating the images in both bands, calculating the value of $\alpha_{\rm{mm}}$ and its standard error in each pixel, and radially averaging these values. Full maps of $\alpha_{\rm{mm}}$ and $\sigma(\alpha_{\rm{mm}})$ are included in Appendix~\ref{app:alphaims}, but here only the radially averaged profiles will be used.
	
	The uncertainty in $\alpha_{\rm{mm}}$ at any point is determined primarily by two components: first, the noise properties of the images in both bands, and second, the absolute flux calibration errors in the images.
	 
	\refedit{The image noise consists of the random noise term to $\alpha_{\rm{mm}}$ which affects each beam separately, but is also affected by residual phase noise. This is particularly relevant for the Band 7 data. Figure~\ref{fig:alphamap} shows that the effect of this type of error does propagate into the $\alpha_{\rm{mm}}$-map. However, the result is that the radial spectral index profile becomes much noisier in those low $S/N$ regions where the azimuthal asymmetries are located, which happens at distances beyond $\sim 1''(150$\,AU). Therefore, we restrict the range of radii over which we discuss the spectral index variations to within this region.}
	
	\refedit{In contrast to the random noise, t}he absolute flux calibration error adds a constant at all positions. Since our data have been self-calibrated, and the rms phase noise on the shortest baselines is less than $15\%$, assuming a $10\%$ flux calibration error is appropriate in both bands~(\citet{almaman,macgregor17}; see also \S~\ref{sec:obsred}). This uncertainty leads to an 0.42 error on $\alpha_{\rm{mm}}$. A third source of error, specific to the radial profile of the spectral index, is the impact of our self-calibration models on the resulting image properties. It is difficult to quantify the result of slightly different source models. However, in general, their main influence should be constrained to changes in the steepness of $\alpha_{\rm{mm}}(R)$ in the bright inner disk region (the inner $38$\,AU), where the self-calibration's impact is the largest, and --- partly --- in extra noise in the radial profile.
	
	The profile of the spectral index $\alpha_{\rm{mm}}(R)$, shown in Figure~\ref{fig:alphar} shows several key features: a low spectral index in the inner disk, and a peak and subsequent drop at the location of the first gap identified in \S~\ref{sec:uvplane}. Here we explore these features in more detail.
	
	\subsubsection{\refedit{A cold inner disk}}
	In the inner beam of our observations (dominated by the bright central component with characteristic radius $38\,$AU in the model fit in \S~\ref{sec:uvplane}, see Figure~\ref{fig:annotated}), the spectral index appears to be consistent with $\alpha_{\rm{mm}} = 2.09 \pm 0.03 $, increasing outward. The dominant source of error in this region is the flux calibration uncertainty in both bands, resulting in an absolute uncertainty in $\alpha_{\rm{mm}}$ over the entire image of 0.42.
	
	\refedit{If we assume the emission is in the Rayleigh-Jeans regime, then $\beta \lesssim 0.5$ - a value that can only be reached if the particles have grown significantly, possibly down to gray ($\beta = 0$) opacity indicating grains with sizes much larger than the wavelengths of these observations. On the other hand, values for $\alpha_{\rm{mm}} \leq 2$ can be explained if the emission in this part of the disk is optically thick, or if the Rayleigh-Jeans approximation does not apply. In both of these cases, the intensity of the central part of the disk becomes sensitive to the temperature of the dust. The observed Band 6 peak intensity of 0.022\,Jy\,beam$^{-1}$ in the central 0.25$''$ beam corresponds to a brightness temperature of only 12.9\,K if we assume the emission is optically thick. This suggests that the disk midplane could be cold enough for CO to be frozen out within a $\sim 20$\,AU radius.}
	
	To test the hypothesis of an optically thick inner disk, an independent dust temperature estimate is necessary. Previously, a temperature profile for V1094 Sco has been given by~\citet{tsukagoshi11}, who fit a power-law temperature structure to the disk's spectral energy distribution (SED). In their best-fit model, $T(R) = 99.3\,\rm{K} \left( \frac{R}{1\,\rm{AU}} \right)^{-0.68}$, giving a temperature of 13\,K at 20\,AU.	An additional check of a plausible range of midplane temperatures was done with the physical-chemical disk modeling code~\texttt{DALI}~\citep{dali1, dali2}, which was used to perform dust radiative transfer calculations for a set of 45 disk models to obtain their dust temperatures\refedit{, raytrace their emission, and image the data with the same $0.25''$ beam used for the observations.} These models were based on a parametric surface density structure and dust properties with values in a range that produces a near- to mid-infrared SED similar to that of V1094 Sco. The parameter ranges and sampling for these models are given in Appendix~\ref{app:sed}, Table~\ref{tab:sedpar}.
	
	In both approaches, the underlying disk density structure is smooth and does not contain rings or gaps, and both therefore have radially decreasing midplane temperature profiles. In the~\citet{tsukagoshi11} model, the midplane temperature at $20$\,AU is $13$\,K. In the grid of~\texttt{DALI} models, we find midplane temperatures of $20 \pm 4$\,K at this radius\refedit{. The resulting raytraced images of the disk show peak brightness temperatures between 17 and 10\,K and a median temperature of $12.7$\,K, very close to the observed peak temperature. This shows that our observations are consistent with a cold inner disk for V1094 Sco.}
	
	An \refedit{important} consequence of the low midplane temperatures predicted by these models is that the midplane CO snowline, which is expected to be around 20\,K for pure CO ice~\citep{oberg05}, is not or in the most optimistic cases barely resolved. The gaps and rings are located far outside the inner $20$\,AU, meaning that another explanation for the presence of these features than the influence of a snowline is needed.
	
	\refedit{If the mm emission in this region is indeed optically thick, this implies that additional substructures could be present in the inner disk, not revealed by these data. High-angular resolution $\geq 3$\,mm data are needed to probe this region, and will also allow for a large reduction in systematic uncertainty in $\alpha_{\rm{mm}}$ across the disk.}

	\subsubsection{Radial spectral index variations in the first gap/ring-pair}
	The high $S/N$ achieved in the image plane in both bands allows us to describe the radial variations of $\alpha_{\rm{mm}}$ out to a radius of $\sim 150$\,AU. Crucially, this region contains the first gap/ring pair in the disk, allowing us to place some constraints on the grain properties in these features.
	
\begin{figure}[!htb]
\begin{center}
		\includegraphics[width=0.5\textwidth]{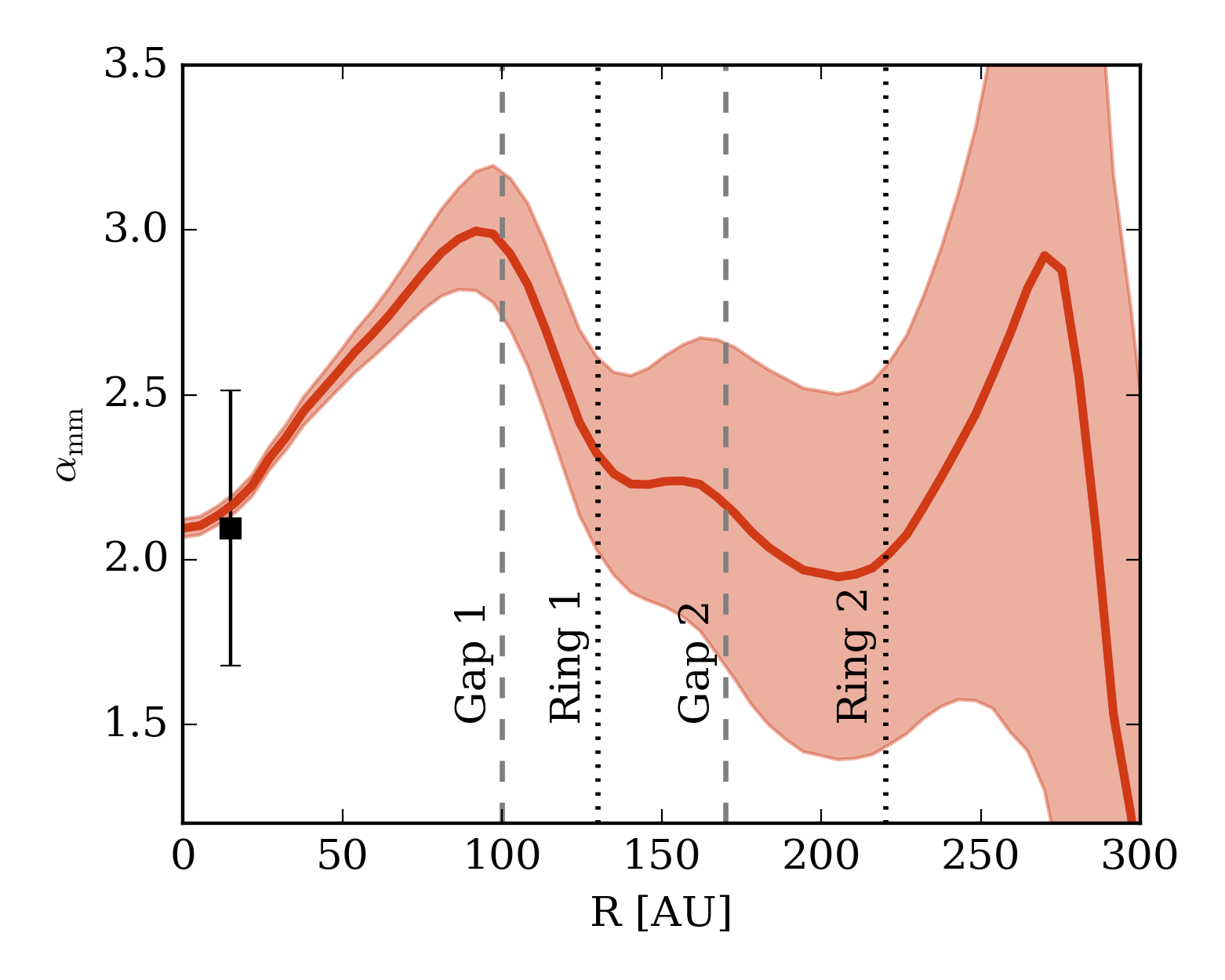}
		\caption{$\alpha_{\rm{mm}}(R)$ (red) for V1094 Sco. $1\sigma$ errors from image noise are in light red, the $10\%$ flux calibration error is shown by the black error bar. The gaps are indicated with dashed gray lines, dotted black lines correspond to the bright rings.}
		\label{fig:alphar}
\end{center}
\end{figure}

	As can be seen from Figure~\ref{fig:alphar}, $\alpha_{\rm{mm}}$ increases from $2.1 \pm 0.03$ in the inner disk, by $\sim 0.9$, towards $3.0 \pm 0.2$ in the first gap at $100$\,AU. Subsequently, the spectral index drops by $\sim 0.7$ to $2.3 \pm 0.3$ in the first ring at $130$\,AU. Some tentative hints of a subsequent increase in $\alpha_{\rm{mm}}$ can be seen towards the location of the second gap, but are very small compared to the (large) increase in image noise at those radii, and will therefore not be further discussed.
	
	Due to the previously mentioned absolute flux calibration error, two possible interpretations for the behavior of $\alpha_{\rm{mm}}$ between $20 - 130$\,AU can be found. On the one hand, the inner region may be optically thick. In this scenario, the material in the gap is optically thin with a $\beta \sim 1$, consistent with grain growth up to mm sizes in this region~\citep{draine06}, and the material in the first ring may be optically thick as well, or, interestingly, have a very low $\beta (\sim 0.3)$ due to a population dominated by grains with sizes larger than a few cm\refedit{, even if the emission is not in the Rayleigh-Jeans regime}.
	
	Alternatively, the entire disk could be optically thin, so that $\beta$ becomes even higher (up to $\sim 1.4$) in the gap, requiring grains in the this part of the disk to be small ($\mu$m-sized), while larger grain sizes are found in the rings and central disk. Indeed, in simulations, local pressure maxima in the disk can reproduce both rings and gaps in disks as well as this type of spectral index profile, as shown in, amongst others,~\citet{pinilla12}.
	
	Intriguingly, similar radial spectral index variations have so far also been identified observationally in both TW Hya and HL Tau~\citep{huang18,liu17, hltau}, suggesting that this behavior occurs more often in rings and gaps in protoplanetary disks. A decreasing $\beta$ with radius is also present in HD 163296, although these observations lack the $S/N$ to resolve the rings~\citep{guidi16}.

\section{Discussion}

	\subsection{Comparing multi-ringed disks}
	\label{sec:ringcomp}
	Besides studying the detailed properties of a single object, the comparison of V1094 Sco with other multi-ringed disks, as well as with other disks in the Lupus star forming region, can provide insight into the mechanisms that drive this peculiar emission morphology.
	
	Despite the small size of the sample of multi-ringed disks found in
the literature, there is quite a large spread in the temperatures and
luminosities of the stars around which they are found. For example, HD
163296 has an effective temperature of $9333$\,K and a luminosity of
$36$\,$L_{\odot}$~\citep{natta04} while V1094 Sco is both much cooler
at $T_{\rm{eff}} = 4205$\,K and an order of magnitude fainter with
$L_{\star} = 1.95$\,$L_{\odot}$~\citep{alcala17}.

\begin{figure*}[!htb]
\begin{center}
		\includegraphics[width=\textwidth]{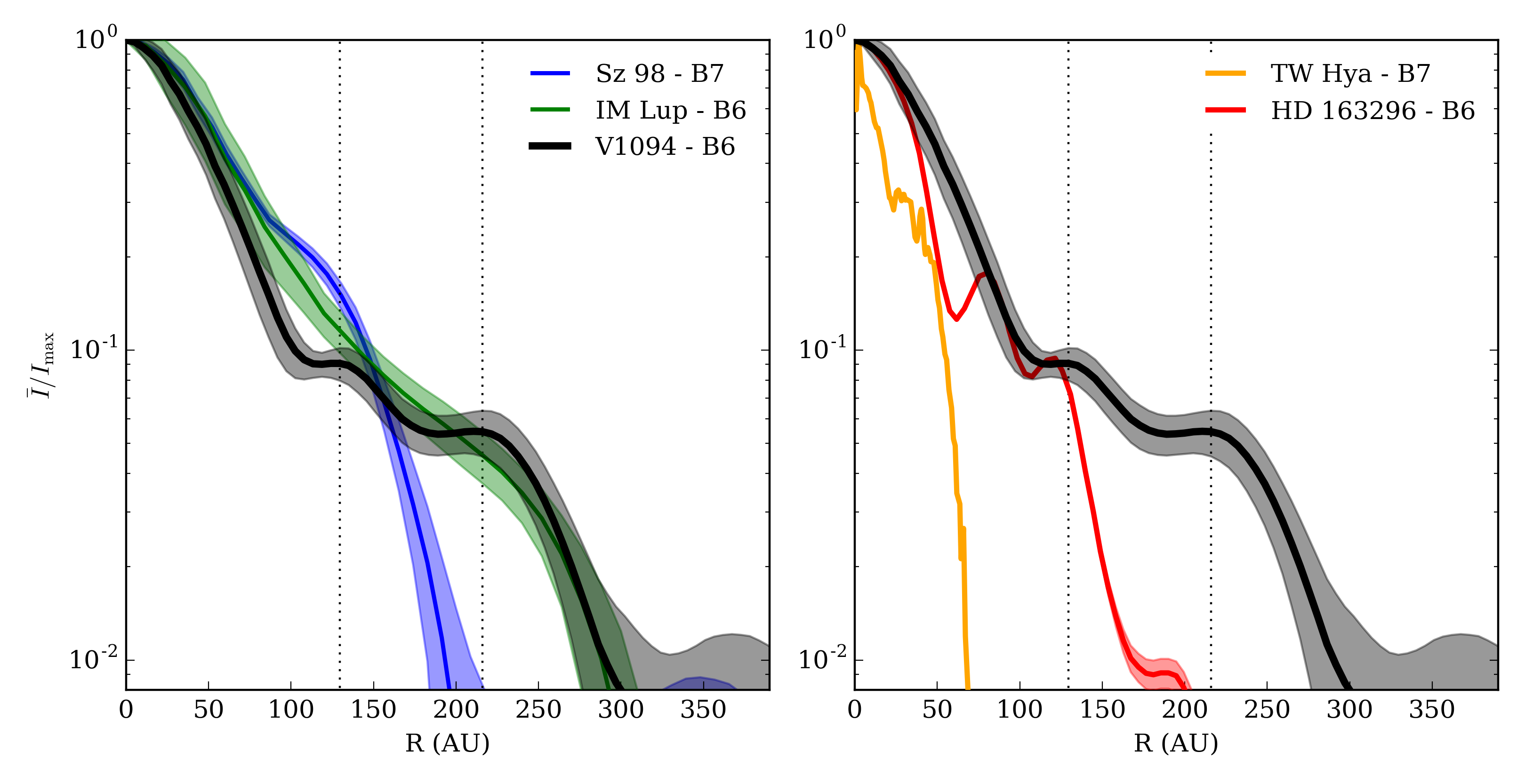}
		\caption{The normalized radial intensity profile of V1094 Sco compared to two large Lupus disks (IM Lup and Sz 98;~\citealt{ansdell16, cleeves16}) (left panel) and to two multi-ringed disks (TW Hya and HD 163296;~\citealt{twhya,isella16}). The locations of the rings of V1094 Sco are indicated with vertical dotted lines. For all Lupus sources, a 150\,pc distance is assumed. \refedit{The effective resolution of the Sz 98 and V1094 Sco disks is $38$\,AU; for IM Lup, $65$\,AU; for TW Hya, $1.6$\,AU; for HD 163296, $27$\,AU.}}
		\label{fig:radcuts}
\end{center}
\end{figure*}

	To facilitate the comparison between other disks in the literature, Figure~\ref{fig:radcuts} presents the normalized radial intensity profile of V1094 Sco compared to several other protoplanetary disks.
The left panel compares the V1094 disk with two of the other largest and brightest Lupus disks, both around relatively cool stars: IM Lup ($T_{\rm{eff}} = 4350$\,K) and Sz 98 ($T_{\rm{eff}} = 4060$\,K)~\citep{ansdell16,cleeves16,alcala17}. The right panel compares disk around V1094 with that of TW Hya, an old, massive, multi-ringed disk around a cool star with $T_{\rm{eff}} = 3800$~K \citep{sokal17}, and the previously mentioned Herbig Ae star HD 163296.  All radial distances of the Lupus disks have been calculated assuming a uniform 150 pc distance (as opposed to the 200\,pc distance, used in~\citealt{ansdell16,ansdell18}).

	Several conclusions can be drawn from Figure~\ref{fig:radcuts}. First, it is clear that the continuum emission from V1094 Sco is similar in extent to that of IM Lup~\citep{cleeves16}, with the caveat that the IM Lup observations shown here are with a larger beam width of $0.4''$, and the outer radius might thus be smaller. This disk also seems to show a continuum break, at similar radii of $\sim 100$\,AU and relative intensity to V1094 Sco. Such a break is also seen in Sz 98, albeit at a higher intensity relative to the peak flux and a smaller radius ($\sim 80$\,AU). Intriguingly,~\citet{tazzari17} found ring-like residuals for this disk when fitting it with a smooth decreasing surface density profile, implying there may be additional structure present in this object. The older TW Hya disk, in contrast, is much more compact than the Lupus disks compared here. This source also shows substructure down to much smaller scales, which would not be resolved with the beam used for the other sources, demonstrating that rings in protoplanetary disks are not necessarily limited to the most radially extended objects.

	It is illustrative to compare the observed intensity profile of V1094
with that HD 163296 as derived by~\citet{isella16}. While HD 163296 is
more compact than V1094 Sco, the two disks share remarkably similar
intensity profiles, with a central core and more extended outer
region, which hosts two rings. The more compact central core of the HD 163296 disk is at least in part due to the smaller beam of the~\citet{isella16} observations ($0.2''$ at a distance of $122$\,pc).

	The difference in ring locations around both stars is significant, since it provides an additional, independent indication that snow lines are an unlikely explanation for the rings in V1094 Sco: in the hotter HD 163296 system, the rings are located further inward. This contrasts with the result in~\citet{zhang16}, who found an overlap in the (inner) ring and gap of HD163296 and the CO snowline, as well as coinciding rings and CO snowlines in the T Tauri stars in their sample (HL Tau and TW Hya). However, both of these T Tauri disks are much more compact than the V1094 Sco disk.
	
	An alternative explanation for the formation of rings, the presence of a magnetohydrodynamic dead zone, also does not appear to be sufficient. Although the models presented by~\citet{flock15}, for instance, show a ring and gap at the outer edge of the dead zone, their model does not lead to a double ring structure, such as in these disks. However, the total (gas$+$dust) column density where the~\citet{flock15} model predicts a gap and ring is similar (at $\sim 3$\,g\,cm$^{-2}$) to that of the best-fit V1094 Sco model of~\citet{tsukagoshi11} at 100\,AU, indicating that one of the rings could be linked to the dead zone edge. Alternatively, the change in turbulence inside the dead zone may be an explanation for the bright inner region of the disk and the `fried-egg' morphology of the emission. The volatile snow lines, located inside 20\,AU, may contribute to the bright inner core region, as per~\citet{banzatti15}.
	
	A (secular) gravitational instability (GI) or the effect of planets are both plausible mechanisms for ring formation at large radii. Secular GI effects can form rings at 100\,AU scales around Solar-like stars~\citep{takahashi14}. Likewise, it is well established that one (or more) previously formed planets of sufficient mass can form gaps and rings of dust in protoplanetary disks~\citep{papaloizou84}; however, the nature of the planets (their mass, and the number of planets per gap) is difficult to constrain based on these data alone. Observations of V1094 Sco by the {\it SEEDS} survey have not led to the detection of a companion~\citep{uyama17}; however, the median sensitivity of those observations is between $4 - 10$\,$M_J$ in the $100-300$\,AU region. This means that even a quite massive planet could easily escape detection using current instruments. Alternatively, perhaps several lower-mass planets are responsible for one or both of the observed gaps.
	
	\subsection{The occurrence rate of large dust disks}
	Since our survey of disks in Lupus is --- with the addition of this sample from the Lupus completion survey (see \S~\ref{sec:obsred}) --- complete, it is now for the first time possible to place concrete upper limits on the occurrence rate of bright, radially extended disks.~\refedit{Here, we look at the radii of disks based on their mm-continuum emission. Although $^{12}$CO emission is often more extended than the continuum,~\citep{ansdell18} show that the average ratio $R{\rm{gas}}/R_{\rm{dust}}$ is within a factor $1.5 - 3$ in Lupus.} In total, there are $95$ Class-II sources in the Lupus clouds. Of these disks, IM Lup and V1094 Sco are by far the most extended with radii up to $\sim 300$\,AU, assuming a 150\,pc distance, with the next largest disk at a radius of $\sim160$\,AU. This implies an occurrence rate of $2.1 \pm 1.5 \%$ for these largest disks.

	Given the high surface densities at large radii of disks like V1094 Sco and IM Lup, and the finding of rings in V1094 Sco that could be sculpted by planet-disk interactions, it is interesting to compare the statistics of large disks to the number of massive planets at large orbital radii.~\citet{bowler16} find an occurrence rate of $<2.1 \%$ for planets in orbits between $100 - 1000$\,AU with masses of $5 - 13\,M_J$, around $5-300$\,Myr-old stars across all spectral types.
	
	In contrast, the number of full (non-transitional) disks in Lupus where continuum emission was detected beyond $100$\,AU is significantly larger: $8 \pm 3.0 \%$ of the full population~\citep{vandermarel18}. Since we might not detect disks with fainter continuum emission at large radii, this is a lower limit to the actual number of these disks.
	
	If we also consider the Lupus transition disk population, the number of disks with continuum emission beyond 100\,AU further increases, to $18 \pm 4.3 \%$. However, these transition disks are consistent with clearing by planets interior to the gap radius, which is $<100$\,AU for the entire sample~\citep{vandermarel18}. Thus, the transition disk population is not necessarily comparable to the statistics of massive exoplanets in the $100-1000$\,AU interval.
	
	The exoplanet occurrence rates discussed here are, in either case, much smaller than the number of disks with significant amounts of continuum emission at similar radii. However, the statistics of the extrasolar planet population at these orbital radii are very incomplete, especially for lower-mass planets ($<5\,M_J$) around fainter stars. Thus, it is possible that the discrepancy in the occurrence rate of planets at large orbital radii is due to the disks forming mostly lower-mass planets. Indeed, no particularly massive planets are needed to open detectable gaps at large orbital distances, since the Hill sphere radius increases linearly in orbital distance but only as the third root of mass. Both~\citet{isella16} and~\citet{dipierro18}, for instance, find planet masses below $5\,M_J$ are sufficient to form gaps in HD 163296 and Elias 2-24 respectively. Alternatively, the large difference in planet detection statistics could be due to significant planet migration moving most of the planets that formed at large distances from their stars to smaller orbits.
	
	Apart from comparing the population of large disks directly to the exoplanet population, relating large primordial disks to the entire Lupus transition disk sample is also informative. Interestingly, although transition disks with large cavities make up $\gtrsim 11 \pm 3.7 \%$ of the disks in Lupus~\citep{vandermarel18}, they are strongly overrepresented in the sample of largest disks: $50\%$ of disks with continuum radii in the upper octile show evidence of central cavities in their continuum emission. This led~\citet{vandermarel18} to propose a two-mode evolutionary sequence for protoplanetary disks. In this view, large primordial disks like V1094 Sco and IM Lup are the progenitors of transition disks, while the majority of lower-mass, more compact disks may evolve through a different pathway, either without a transitional stage or as a scaled-down version of the pathway of the large disks. The massive primordial disks may trap dust by some means (possibly through planet-disk interactions, or by some other mechanism), which prevents its inward drift. This leads to the formation of rings and, later, the emergence of a large, possibly multi-ringed, transition disk with a large dust cavity.
	
	In the context of this hypothesis, it is possible to interpret the ratio of large primordial disks to transition disks as an indication of the speed at which this evolution occurs. Since half the disks larger than $120$\,AU are transition disks at the $1 - 3$\,Myr age of Lupus, this process should take place on $\sim 1$\,Myr timescales. Importantly, this means we have a testable prediction for the occurrence rate of large primordial disks in other star forming regions: they should be rarer in more evolved regions, and more common in younger systems, if the absolute rate of evolution across regions is the same. In terms of disk mass, too, it is then expected that the disk mass distribution in more evolved regions is dominated by a small number of transition disks, while younger regions should have an upper mass distribution with mostly large primordial disks. Although statistics across other regions at similar sensitivity and resolution are still lacking, this framework is at least consistent with the observation of several large, multi-ringed systems in very young systems: HL Tau, Elias 2-24 and AS 209 are all at most 1\,Myr old~\citep{hltau,andrews10,dipierro18,natta06,fedele17b}. However, such young systems tend to be brighter, leading to an observational bias for high-resolution imaging campaigns towards these disks, which in turn results in their overrepresentation in the already small sample of multi-ringed disks.

\section{Conclusions}
	In this work the nature of V1094 Sco as a giant, multi-ringed protoplanetary disk is identified for the first time, based \refedit{on }data from the Lupus survey of protoplanetary disks. Analyzing its structure and comparing this disk to the rest of the Lupus disk population allows us to exclude some pathways for ring formation, and provides a possible link to a general disk formation picture.
	\begin{itemize}
		\item Continuum emission from V1094 Sco extends out to 300\,AU from the central star, making it larger than the other Lupus disks by a factor $\sim 5$, but similar in size to the well-studied IM Lup disk.
		\item Two bright rings are identified at 130 and 220\,AU, separated by gaps at 100 and 170\,AU, based on a u,v-plane fit of the intensity profile.
		\item Disks with continuum radii of this size ($> 200$\,AU) are rare: only $\sim 2\%$ of the Class II objects in Lupus are giant disks similar to V1094 Sco.
		\item Radial variations in $\alpha_{\rm{mm}}$ are consistent with optically thick emission in the bright disk core. The spectral index increases at the location of the first gap, and subsequently decreases again towards the first ring, possibly as a result of efficient grain growth in this ring. 
		\item Based on the temperature of the optically thick emission and on independent disk models, the midplane dust temperature at the ring locations is too low for snow lines to be a valid explanation, while a dead zone would only be expected to cause one ring. However, gravitational instabilities or the presence of planets up to $\sim 10$\,$M_J$ in the gaps cannot be excluded.
		\item The number of primordial disks with continuum emission at radii $> 100$\,AU in Lupus seems to be much larger than the occurrence rate of giant planets with $5 - 13\,M_J$ between $100-1000$\,AU. The large disk/transition disk ratio may be a probe of disk evolution, if large multi-ringed disks evolve into (multi-ringed) transition disks with large cavities on $\sim$Myr timescales.
	\end{itemize}
	
\begin{acknowledgements}
Astrochemistry in Leiden is supported by the European Union A-ERC grant 291141 CHEMPLAN, by the Netherlands Research School for Astronomy (NOVA), and by a Royal Netherlands Academy of Arts and Sciences (KNAW) professor prize. This paper makes use of the following ALMA data: ADS/JAO.ALMA\#2016.1.01239.S. ALMA is a partnership of ESO (representing its member states) NSF (USA) and NINS (Japan) together with NRC (Canada) MOST and ASIAA (Taiwan), and KASI (Republic of Korea), in cooperation with the Republic of Chile. The Joint ALMA Observatory is operated by ESO, AUI/NRAO and NAOJ. The authors acknowledge support by Allegro, the European ALMA Regional Center node in The Netherlands.\end{acknowledgements}

%%% Bibliography goes here %%%
\bibliographystyle{aa}
\bibliography{V1094_bib}

\begin{appendix}

\renewcommand\thefigure{\thesection.\arabic{figure}}
\section{Spectral index map of V1094 Sco}
\setcounter{figure}{0}
\label{app:alphaims}

\begin{figure*}[!h]
\begin{center}
		\includegraphics[width=1\textwidth]{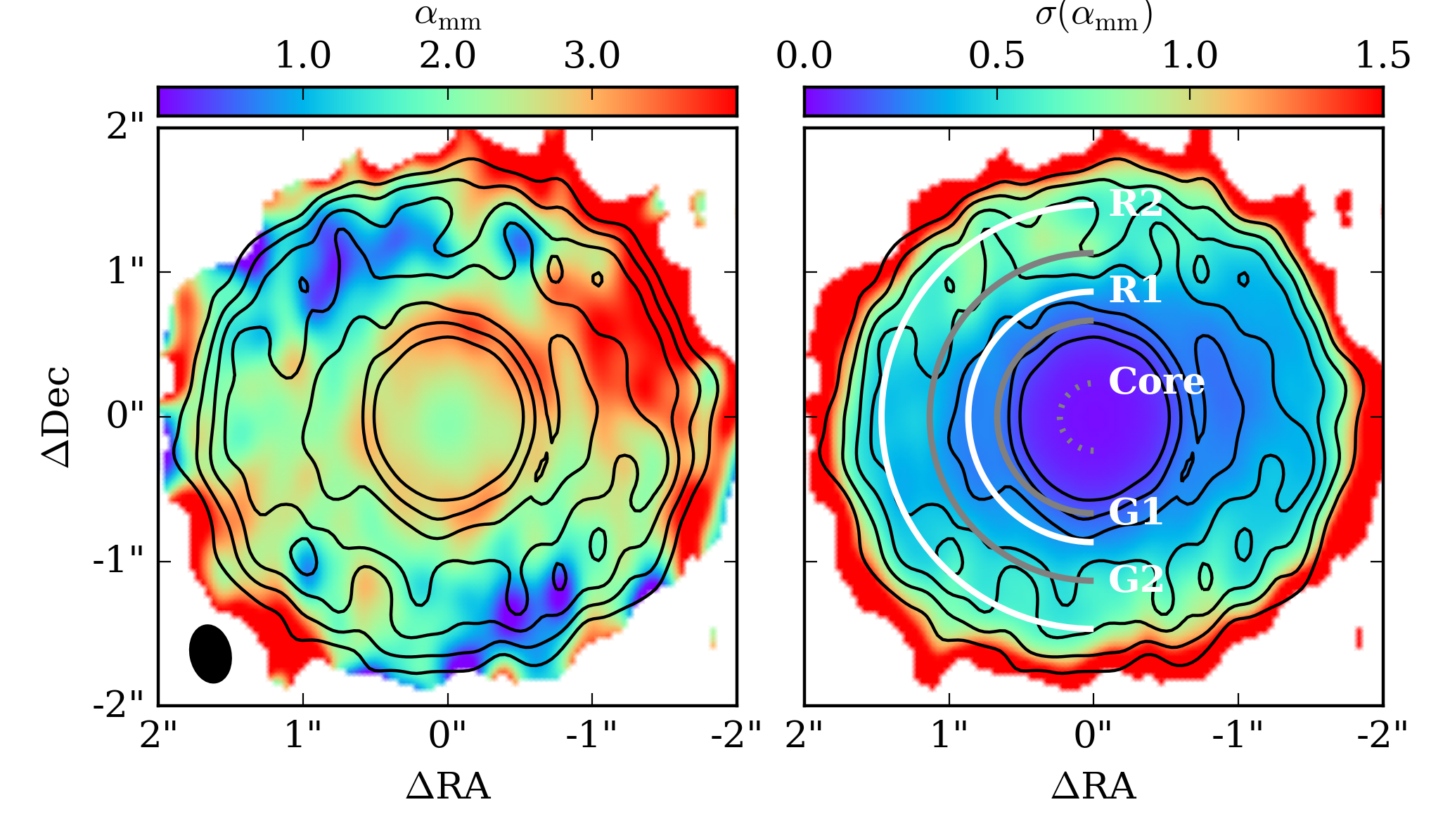}
		\caption{\textit{Left}: $\alpha_{\rm{mm}}$, deprojected and derotated. \textit{Right}: Standard error on $\alpha_{\rm{mm}}$, not including the $\pm 0.42$ offset due to flux calibration errors. The Band 6 data are overplotted in black, with contours at 3, 5, 7.5, 10, 15, 20 and 30$\sigma$. The locations of the rings (R1 and R2), gaps (G1 and G2) and the bright core are indicated with white and gray arcs.} 
		\label{fig:alphamap}
\end{center}
\end{figure*}

Figure~\ref{fig:alphamap} shows the full maps of $\alpha_{\rm{mm}}$ and $\sigma(\alpha_{\rm{mm}})$ that were used to construct the radial $\alpha_{\rm{mm}}$ profile discussed in Section~\ref{sec:alpha}. Pixels in both Band 6 and Band 7 with $S/N < 2.5$ were masked.

\section{Midplane temperatures of V1094 Sco disk models}
	\setcounter{figure}{0}
	\label{app:sed}
	In Table~\ref{tab:sedpar}, the parameters chosen to for our \texttt{DALI} mini-grid of V1094 Sco-like disks are listed. \texttt{DALI}~\citep{dali1,dali2} is a physical-chemical code that was used to perform dust radiative transfer calculations for the range of dust distributions, and to obtain the dust temperatures and raytrace their emission.  
	
	The disk models themselves are based on the parametric disk density structure introduced in~\citet{andrews11}, here truncated at a radius $R_c$, so that $\Sigma(R) = \Sigma_c \left(\frac{R}{R_c}\right)^{-\gamma} \exp \left( (R/R_c)^{2-\gamma}\right)$. $R_c$ was set to 250\,AU based on the Band 6 continuum image, and we fix $\gamma$ to $0.6$. For a full disk, the key parameters are then the total mass (gas + dust) $M_d$, the scale height $h_c$ (the standard deviation of the vertical Gaussian distribution of the dust), flaring angle $\psi$ (defined so that $h(R) = h_c (R/R_c)^{\psi}$), the large grain fraction $f_{ls}$, and the relative settling factor $\chi$, which sets the large grain scale height by $h_{c\,\rm{large}} = \chi h_c$. Large grains have sizes up to $1000\,\mu$m, while small grains do not exceed $1\,\mu$m in size. The total mass was fixed at $0.1\,M_{\odot}$, reproducing the Band 6 continuum flux assuming a gas to dust ratio of 100.

\begin{table}[hb]
\caption{Parameter ranges and number of samples for the disk models}
\label{tab:sedpar}
	\centering
	\begin{tabular}{l l l l l}
	\hline \hline
	Parameter & minimum & maximum & $N_{\rm{samp}}$ \\
	\hline
	Scale height $h_c$ & 0.02 & 0.1 & 4 \\
	Flaring angle $\psi$ & 0.1 & 0.3 & 3 \\
	Large grain fraction $f_{ls}$ & 0.85 & 0.99 & 2 \\
	Settling parameter $\chi$ & 0.2 & 1.0 & 2 \\
	\hline
	\end{tabular}
\end{table}

\begin{table}[]
\caption{\refedit{Stellar properties used for the disk models}}
\label{tab:starpar}
	\centering
	\begin{tabular}{l l l l l l}
	\hline \hline
	%Parameter & minimum & maximum & $N_{\rm{samp}}$ \\
	$M_{\star}$\tablefootmark{a} & $R_{\star}$   & $L_{star}$    & $T_{\rm{eff}}$ & SpT & $d$ \\
	$[M_{\odot}]$                & $[R_{\odot}]$ & $[L_{\odot}]$ & $[\rm{K}]$     &     & pc  \\
	\hline
	0.92          & 1.9           & 1.7           & 4205           & K6  & 150 \\
	\hline
	\end{tabular}
	\tablefoot{
	\refedit{Values derived from~\citet{alcala17} assuming a 150\,pc distance to the disk.}\\
	\tablefoottext{a}{\refedit{Mass derived with the \citet{siess00} model grid.}}
	}
\end{table}

\begin{figure}[!htb]
\begin{center}
		\includegraphics[width=0.5\textwidth]{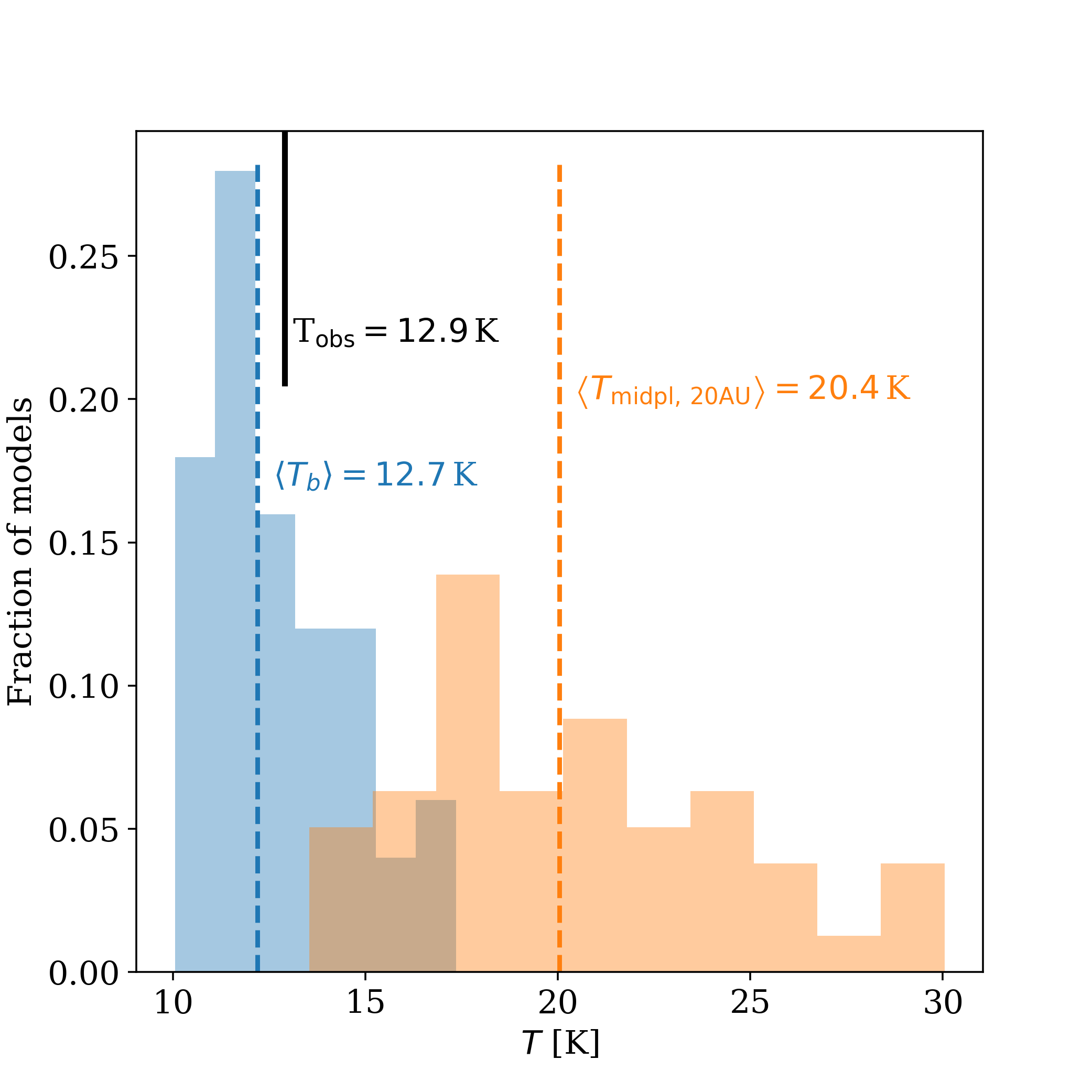}
		\caption{\refedit{\texttt{DALI} model midplane temperatures at 20\,AU (orange) and Band 6 peak brightness temperatures (blue) after convolution with an $0.25''$ beam for a grid of disks. The black line indicates the observed peak brightness temperature of V1094 Sco. For pure CO ice in typical disk midplane conditions, the freeze-out temperature is 20 K ({\"O}berg et al. 2005).}}
		\label{fig:sedpar}
\end{center}
\end{figure}

\end{appendix}

\end{document}